**Integrating network pharmacology, metabolomics, and gut microbiota analysis to explore the effects of Jinhong tablets on chronic superficial gastritis**

Lihao Xiao[b,c,#], Tingyu Zhang[a,#], Yun Liu[a], Chayanis Sutcharitchan[a], Qingyuan Liu[a], Xiaoxue Fan[c], Jian Feng[c], Huifang Gao[c], Tong Zhang[b,*], Shao Li[a,*]

[a] Institute for TCM-X, Department of Automation, Tsinghua University, 100084 Beijing, China

[b] School of Pharmacy, Shanghai University of Traditional Chinese Medicine, 201203 Shanghai, China

[c] State Key Laboratory on Technologies for Chinese Medicine Pharmaceutical Process Control and Intelligent Manufacture, Jiangsu Kanion Pharmaceutical Co.,Ltd., 210000 Nanjing, Jiangsu, China.

#: These authors contributed equally.
*: Corresponding author:
Shao Li, Institute for TCM-X, Department of Automation, Tsinghua University, 100084, Beijing, China
tel: +86 10 62797035; fax: +86 10 62786911.
E-mail address: shaoli@mail.tsinghua.edu.cn
Tong Zhang, School of Pharmacy, Shanghai University of Traditional Chinese Medicine, 201203, Shanghai, China
E-mail address: zhangtongshutcm@hotmail.com

**Abstract**

Background: Chronic gastritis (CG) significantly impacts patients' quality of life and can progress to more severe gastric conditions. In China, Traditional Chinese Medicine (TCM) has been widely applied for its holistic efficacy in treating chronic superficial gastritis (CSG), including formulas like Jinhong Tablets (JHT), known for their anti-inflammatory effects. However, the mechanism of action of JHT in treating CSG still require further clarification.

Purpose: This study aimed to elucidate the mechanism by which JHT alleviates CSG, integrating network pharmacology, untargeted metabolomics, and gut microbiota analyses.

Methods: The CSG rat model was established, and treatment effects were assessed via Hematoxylin and eosin (H&E) staining. Target profiles of JHT's components and holistic targets of JHT were obtained. Enrichment analyses were performed on holistic targets and a multi-layer biomolecular network of JHT was established. The study also analyzed rat plasma for differential metabolites through untargeted metabolomics and evaluated the diversity and composition of gut microbiota in fecal and cecal contents samples using 16S rRNA sequencing.

Results: JHT effectively reduced gastric inflammation in CSG rats. Network pharmacology indicated that diverse metabolic processes including lipid metabolism and nitric oxide metabolism play pivotal roles in the therapeutic effects of JHT on CSG. Metabolomics analysis identified differential metabolites including betaine helping enrich gut microbiota, phospholipids and citrulline indicating the severity of CSG. The pathway enrichment of differential metabolites validated the results of network pharmacology and suggested that the mechanism of action was associated with the gut microbiota. Through gut microbiota analyses, it was discovered that JHT could augment the gut microbiota by enhancing the abundance of betaine. Additionally, JHT was shown to boost the production of short-chain fatty acid (SCFA) by increasing the abundances of Faecalibaculum and Bifidobacterium, consequently alleviating gastric inflammation in CSG.

Conclusion: Our study revealed that JHT alleviated CSG through diverse metabolic processes including lipid and energy metabolism. Metabolites such as betaine, along with gut microbiota including Faecalibaculum and Bifidobacterium, play crucial roles in the therapeutic interventions. Our findings support the therapeutic potential of JHT and contribute to a deeper understanding of the role of TCM in the treatment of chronic superficial gastritis.

**Abbreviations:**

CG: chronic gastritis; CSG: chronic superficial gastritis; IM: intestinal metaplasia; CAG: chronic atrophic gastritis; JHT: Jinhong tablets; TCM: traditional Chinese medicine; MT: Melia toosendan; CY: Corydalis yanhusuo; VS: Vladimiria souliei; ID: Illicium dunnianum; UPLC-Q/TOF-MS: ultra-performance liquid chromatography coupled with quadrupole time-of-flight tandem mass spectrometry; WSS: within sum of square; KEGG: Kyoto Encyclopedia of Genes and Genomes; GO: Gene Ontology; H&E: Hematoxylin and eosin; QC: quality control; UHPLC-Orbitrap-MS/MS: ultra-high-performance liquid chromatography-orbitrap mass spectrometry; HESI: high-resolution electrospray ionization; HCD: higher-energy collisional dissociation; PCA: principal component analysis; OTU: operational taxonomic unit; PLS-DA: partial least squares discriminant analysis; LDA: linear discriminant analysis; C: control; M: model; D: low-dose JHT; G: high-dose JHT; TLRs: Toll-like receptors; SGJY: Shugan Jieyu; LQHX: Liqi Huoxie; HWZT: Hewei Zhitong; PS: phosphatidylserine; PE: phosphatidylethanolamine; SCFA: short-chain fatty acid; GABA: γ-Aminobutyric acid; NO: nitric oxide; NOS: nitric oxide synthase; iNOS: inducible nitric oxide synthase

# Introduction

Chronic gastritis (CG) refers to persistent inflammation primarily affecting the gastric mucosa, often manifesting symptoms such as epigastric pain, bloating, indigestion, nausea, vomiting, belching, and loss of appetite, all of which significantly impact patients' quality of life (Quan et al., 2018). Based on the histopathological characteristics of the gastric mucosa, CG can be categorized into chronic superficial gastritis (CSG) which is also known as chronic non-atrophic gastritis, chronic atrophic gastritis (CAG), and other specific subtypes (Yang et al., 2023). CSG is considered a precursor to more severe gastric diseases, notably progressing to the premalignant stages of CAG and intestinal metaplasia (IM), which may ultimately lead to gastric cancer (Pimentel-Nunes et al., 2019; Zhang et al., 2019). Therefore, early detection and intervention in patients with CSG are critical, not only to relieve symptoms but also to prevent progression toward precancerous stages or even gastric cancer.

Traditional Chinese Medicine (TCM) has been widely applied in the treatment of chronic disease due to its proven efficacy, minimal side effects, and holistic approach (Li et al., 2003). Numerous clinical and mechanistic studies have explored the therapeutic potential of TCM for CG. A multi-center, double-blind, randomized controlled trial on Qirui Weishu capsules demonstrated their effectiveness in relieving gastric pain and reducing gastric mucosal inflammation (Chen et al., 2022). Zhou et al. employed network pharmacology to elucidate the therapeutic mechanisms of Moluodan in the treatment of CAG and validated the findings through experiments (Zhou et al., 2022). Furthermore, Wang et al. integrated omics data and algorithmic analysis to uncover the specific immunomodulatory effects of Weifuchun in treating CAG, which were confirmed through cell line and rat model experiments (Wang et al., 2023b).

Jinhong Tablets (JHT) are composed of four traditional Chinese herbs: *Melia toosendan* (MT, Chuanlianzi), *Corydalis yanhusuo* (CY, Yanhusuo), *Vladimiria souliei* (VS, Chuanmuxiang), and *Illicium dunnianum* (ID, Honghuabajiaoye). According to TCM's theories, these ingredients work synergistically to sooth the liver, relieve depression, regulate Qi, promote blood circulation, and alleviate pain in the stomach. ZHENG, known as TCM syndrome, can help better diagnose and treat diseases (Su et al., 2012). JHT is specifically used for the treatment of CSG which is associated with TCM's liver-stomach disharmony syndrome, characterized by symptoms such as epigastric distention and pain, discomfort spreading to both flanks, acid reflux, belching, white tongue coating, and a wiry pulse.

Recent research has identified the chemical composition of JHT through ultra-performance liquid chromatography coupled with quadrupole time-of-flight tandem mass spectrometry (UPLC-Q/TOF-MS) (Shi et al., 2022). The researchers utilized symptom-guided network pharmacology to reveal the molecular basis of JHT against CSG. Clinical studies have further demonstrated the efficacy of JHT in reducing inflammation and alleviating bile reflux, as confirmed by gastroscopy and biopsy results (Zhang et al., 1998). Additionally, a five-step strategy using UPLC-Q/TOF-MS combined with the Mass Defect Filter technique and MetaboLynx™ software was developed to explore the metabolic profile of JHT in vivo, providing a detailed foundation for further pharmacological research. (Liu et al., 2021). However, the precise mechanisms through which JHT exerts its therapeutic benefits in CSG remain to be fully elucidated.

Network pharmacology, grounded in the principles of systems biology, transcends the limitations of reductionist research that focuses on "single-target" approaches. Its core theory of "network targets" aligns closely with the holistic philosophy and syndrome differentiation-based treatment principles of TCM (Lai et al., 2020; 张 and 李, 2015; 李, 2011). Consequently, network pharmacology has emerged as a cutting-edge methodology in TCM research (Li, 2021a, 2021b). A multi-layered biomolecular network, comprising drugs, targets, pathways, and other molecular components, provides valuable insights into the mechanisms by which drugs intervene in diseases (Huang and Li, 2010). The advent and advancement of multi-omics sequencing technologies, including genomics, proteomics, and metabolomics, provide new insights and foundational support for understanding the mechanisms of herbal medicines and their targets (Wang et al., 2022; Zhang et al., 2023; Wang et al., 2024). The rapid development of some biotechnologies has also provided new tools for the validation of network pharmacology analysis (Tang et al., 2015; Qiao et al., 2018). In addition, some studies have already validated the accuracy and reliability of network pharmacology through high-throughput experiments (Wang et al., 2023; L. Wang et al., 2024; Yang et al., 2024).

The holistic concept represents the core philosophy of TCM, as well as TCM network pharmacology, advocating for a systematic approach to tackle complex diseases. Unlike other network pharmacology analyses, Li et al. introduced a novel TCM network pharmacology paradigm centered on a holistic perspective, subsequently developing a series of innovative methodologies tailored to TCM network pharmacology (Li and Zhang, 2013). Metabolism, as a frontier of systems biology, can reflect the regulatory effects of drugs on the human body through changes in metabolite abundance (Kell and Goodacre, 2014). Metabolomic analysis of peripheral blood serves as a window peeking into the overall metabolic state of the body, aligning seamlessly with the holistic philosophy of network pharmacology, enabling the validation and prediction of results derived from network pharmacology analyses.

The gut microbiota plays a vital role in the human body's immune, metabolic, structural, and nervous systems (Adak and Khan, 2019), exhibiting interconnected regulatory capabilities with organs such as the brain and lungs to maintain systemic homeostasis (Cryan et al., 2019; Dang and Marsland, 2019). Therefore, network pharmacology, untargeted metabolomics, and gut microbiota analysis offer a holistic perspective for elucidating the mechanisms underlying JHT's therapeutic effects against CSG.

This study comprehensively evaluated the pharmacological effects of JHT on CSG rats using an experimental CSG rat model. To elucidate the underlying mechanisms, a multifaceted approach combining network pharmacology, metabolomics, and 16S rRNA sequencing was employed. The findings from network pharmacology were further validated by analyzing metabolic profile changes before and after JHT treatment. Moreover, gut microbiota with significant differential abundance were identified, and their potential interactions with differential metabolites were explored.

## Methods and Materials

### Component identification of JHT

In a previous study, UPLC-Q/TOF-MS was combined with a chemical compound database for the four herbal ingredients of JHT to rapidly characterize the chemical components of this TCM formula (Shi et al., 2022). Additionally, the corresponding information of these components was collected from PubChem database where some components not recorded in database were filtered.

### Target prediction and compounds clustering

DrugCIPHER (Zhao and Li, 2010), a drug target prediction algorithm based on network and compound structural similarity was employed to predict the genome-wide targets of the chemical components in JHT. The top 100 predicted targets, ranked by relevance within the genome-wide target set, were defined as the target profile for each compound. The predicted targets were validated for accuracy using existing literature from the PubMed database and documented records from the PubChem database. The predicted druggable targets of each component were then subjected to an unsupervised hierarchical clustering model. The topological distances for clustering were calculated using Euclidean distance, with the clustering method parameter set to "complete." The optimal number of clusters was determined to be five, based on the elbow plot of within sum of square (WSS).

### Network targets analysis of JHT in the treatment of CSG

Based on the hypothesis from Liang et al.'s study that targets appearing in a greater number of compounds in TCM formulas are likely to play more critical roles, the holistic targets of JHT and each herb were identified (Liang et al., 2014). In addition, enrichment analysis of the holistic targets of JHT was performed on Kyoto Encyclopedia of Genes and Genomes (KEGG) and Gene Ontology (GO) via clusterProfiler v4.11.1. The significantly enriched (P adjust < 0.05) KEGG pathways were identified and filtered to be the CSG key biological pathways. Key biological modules consisting of key biological pathways in the treatment of CSG were defined by KEGG categories and subcategories. Based on predicted holistic targets of each herb and classified key biological modules, a multi-layer biomolecular network was established. The links between herbs and targets show that the genes belong to an herb, while the links between targets and pathways indicate that the genes were significantly enriched in the pathways. The visualization of multi-layer biomolecular network was accomplished by Cytoscape v3.9.1. Additionally, as a supplement, significantly enriched (P adjust < 0.05) GO biological processes were filtered and classified into special categories by biological functions. R packages ggplot2 v3.5.0 and UpSetR v1.4.0 were utilized to implement the remaining visualizations.

**Animal model**

In animal model, male Wistar rats were randomly assigned to four groups: control group (C, n=8), model group (M, n=8), low-dose (70mg/kg) JHT group (D, n=8), and high-dose (280mg/kg) JHT group (G, n=8). The methods for animal modeling and intervention followed the established studies on JHT (B. Wang et al., 2024a), with the comprehensive approach specifically referring to the construction of the CSG rat model.

**Hematoxylin and eosin (H&E) staining and diagnosis of CSG**

The procedure for H&E staining is as follows: (1) sections are sequentially placed in xylene I for 5 minutes, xylene II for 5 minutes, xylene III for 5 minutes, absolute ethanol I for 5 minutes, absolute ethanol II for 5 minutes, 95% ethanol for 5 minutes, 90% ethanol for 5 minutes, 80% ethanol for 5 minutes, 70% ethanol for 5 minutes, 50% ethanol for 5 minutes, and then rinsed with distilled water (Note: In winter, the deparaffinization time should be extended appropriately.); (2) sections are stained in Harris hematoxylin solution for 5 minutes, then rinsed several times in tap water; (3)

sections are differentiated in hydrochloric acid ethanol solution for 1 second; (4) sections are placed in warm water for a few seconds to restore the blue hue (monitor nucleolar clarity under the microscope), followed by rinsing with room temperature distilled water for 1 second; (5) sections are stained with alcoholic eosin solution for 40 seconds, then rinsed with water for 2 seconds; (6) sections are dehydrated through an ethanol gradient and cleared with xylene; (7) sections are mounted with neutral resin. After the above steps, microscopic examination shows the nuclei stained blue, while the cytoplasm and stroma are stained red.

The histopathological grading criteria for gastric tissue are as follows: Grade 0 indicates no inflammatory cell infiltration; Grade 1 represents multiple chronic inflammatory cells observed in the epithelial or basal lamina propria of the gastric mucosa; Grade 2 is characterized by numerous inflammatory cells infiltrating from the epithelial layer to the muscularis mucosae of the gastric mucosa; and Grade 3 shows aggregates of inflammatory cells forming focal clusters in the gastric mucosa. Intermediate grades are assigned by adding 0.5 to the lower grade number.

For each tissue section, three fields of view were observed and scored separately to evaluate the infiltration of inflammatory cells. These three fields of view are: (1) the epithelial layer or basal lamina propria of the gastric mucosa; (2) within the gastric mucosa; (3) from the epithelial layer to the muscularis mucosae and from the muscularis mucosae to the submucosa. Among them, red arrows indicate the inflammatory cells infiltration in the place from the epithelial layer to the muscularis mucosa, while black arrows indicate the inflammatory cells infiltration in the place from the muscularis mucosa to the submucosa. The total inflammation score is calculated as the sum of the scores from these three fields of view. Pairwise differences in inflammation scores between different groups were assessed using t-tests, calculated with the pairwise.t.test function in R, with the p.adjust.method set to Bonferroni for multiple testing correction.

**Sample preparation and detection conditions for untargeted metabolomics**

At the end of the experiment, plasma samples were collected from each animal in control and experimental groups described in the animal model section. Prior to analysis, plasma samples were thawed at room temperature. A precise 50 µL aliquot of plasma was added to 150 µL of methanol containing the internal standard, vortex-mixed for 1 minute, and centrifuged at 12,000 rpm for 10 minutes. The resulting supernatant was then transferred to sample vials for UHPLC-Orbitrap-MS/MS analysis, with all sample processing conducted on ice. Quality control (QC) samples were prepared in the same manner as the afore-mentioned samples, using 50 µL of the mixture of 50 µL of each experimental plasma sample. Before sample analysis, QC samples were analyzed six times consecutively. During sample analysis, one QC sample was analyzed after every ten experimental samples to monitor method consistency.

For the preparation of the internal standard solution, an appropriate amount of 2-chloro-L-phenylalanine reference standard was precisely weighed and placed into a 100 mL volumetric flask. The standard was dissolved and brought to volume with methanol to reach a final concentration of approximately 5 µg/mL, and the solution was stored at 4°C.

The UHPLC-Orbitrap-MS/MS analysis was conducted under specific chromatographic and mass spectrometry conditions. Chromatographically, an Agilent ZORBAX Eclipse Plus-C18 column (2.1 × 100 mm, 1.8 µm) was maintained at 30°C with a flow rate of 0.4 mL/min and an injection volume of 1 µL. The mobile phase consisted of 0.1% formic acid in water (A) and acetonitrile (B) with the following elution gradient: 0–1 min, 2% B; 1–5 min, 2–46% B; 5–8 min, 46–50% B; 8–13 min, 50–60% B; 13–17 min, 60–76% B; 17–25 min, 76–90% B; 25–27 min, 90% B; and 27–29 min, 90–2% B. For mass spectrometry, high-resolution electrospray ionization (HESI) was utilized with both positive and negative ion detection. The ion scan range was set at m/z 100–1500 for MS1 and m/z 50–1000 for MS2. Instrument parameters included a Sheath Gas at 50 Arb, Aux Gas at 10 Arb, Sweep Gas at 1 Arb, Ion Transfer Tube Temperature at 325°C, and Vaporizer Temperature at 350°C. Resolution was set at 60,000 for MS1 and 30,000 for MS2, with an RF Lens value of 70%, and the collision energy for higher-energy collisional dissociation (HCD) was set at 10, 20, and 40 V.

**Untargeted metabolomics analysis**

The LC-MS data from each rat plasma sample were imported into Compound Discovery software, where a series of metabolomics workflows were applied for peak identification, peak filtering, peak alignment, and preliminary identification. The processed data were then exported in Excel format and imported into Mass Profiler Professional (MPP, Agilent Technologies, USA) for sample labeling and grouping. Data processing was conducted based on specific parameters (experiment type: identified and unidentified, species: rat, baseline: median of all samples), including frequency filtering, which retained only data present in at least 80% of samples in any group.

To compare the metabolic profiles among different groups, the standardized rat plasma metabolic data were analyzed in MPP software using unsupervised principal component analysis (PCA) on four groups. The abundance of differential metabolites across these groups was transformed by Z-score normalization and visualized using the GroupHeatmap function of the R packages SCP v0.5.6 and Seurat v5.1.0. Differential metabolites in rat plasma were screened using the Wilcoxon rank-sum test and log Fold Change (logFC). Metabolites showing significant differences ($P < 0.05$) between the M group and C group, as well as between the G group and M group, were identified. The up- or down-regulation of these metabolites was determined based on whether logFC was greater than zero. Compounds with precise molecular weight differences within ±10 ppm were selected, and their MS/MS fragmentation data were used to search for matches in online metabolomics databases, such as HMDB (http://www.hmdb.ca) and mzCloud. Further pathway-based enrichment analysis of these differential metabolites was performed using the Enrichment Analysis function on the MetaboAnalyst website, with the RaMP-DB metabolite set library, which integrates 3,694 metabolite and lipid pathways from sources like KEGG, HMDB, Reactome, and WikiPathways.

**Experimental materials and sample preparation for gut microbiota**

The experimental setup involved various instruments, including NanoDrop 2000 Micro-Volume Spectrophotometer (Thermo Fisher Scientific), Cryogenic Grinder (Shanghai Jingxin Industrial Co., Ltd.), Illumina MiSeq Sequencing System (Illumina Inc.), Bio-Rad T100 Gradient Thermal Cycler (BIO-RAD), BY-R20 High-Speed Refrigerated Centrifuge (Beijing Baiyang Medical Equipment Co., Ltd.), and -80°C Ultra-Low Temperature Freezer (Thermo Fisher Scientific (Suzhou) Instruments Co., Ltd.). Reagents and chemicals used included the Illumina MiSeq platform (Illumina Inc.), FastPfu Polymerase (TransGen Biotech, Beijing), and the Fast DNA SPIN Kit for Soil (Abe Medical Equipment Trade Co., Ltd.). Experimental samples consisted of fecal and cecal contents collected from each animal in control and experimental groups described in the animal model section.

For genomic DNA extraction, the Fast DNA SPIN Kit for Soil was used according to the manufacturer's instructions, and the extracted DNA was quantified with a micro-volume spectrophotometer. PCR amplification targeted the bacterial 16S rRNA gene V3-V4 region, using 338F (5'-ACTCCTACGGGAGGCAGCA-3') and 806R (5'-GGACTACHVGGGTWTCTAAT-3') primers along with high-fidelity polymerase. The amplification products were visualized on a 2% agarose gel and then purified. The purified DNA fragments were then subjected to paired-end sequencing on the Illumina MiSeq Sequencing System, enabling high-throughput sequencing of the 16S rRNA gene region.

**Gut microbiota analysis**

First, the paired-end (PE) reads data obtained from Miseq sequencing were merged based on overlap relationships, followed by quality control and filtering of the sequencing reads. Next, operational taxonomic unit (OTU) clustering analysis and taxonomic classification were performed. After grouping the sequencing samples, partial least squares discriminant analysis (PLS-DA) was conducted at the OTU level. Based on the taxonomic information, the composition and structure of gut microbiota communities across different samples were analyzed at the phylum and genus levels.

Finally, differential analysis of microbiota between groups was carried out. This included the Kruskal-Wallis H test in the mean proportion of microbial taxa, and Lefse multi-level species difference analysis to detect differences across multiple taxonomic levels (phylum, class, order, family, genus, and species). The analysis identified differential species at various levels, with Linear Discriminant Analysis (LDA) scores used to measure the impact of each species on group differences. The higher the LDA score, the greater the influence of that specie on the observed differences. The tools used for analysis are listed in Table 1.

**Results**

**Target prediction and comparison among different types of compounds in JHT**

Based on the JHT chemical components identified in previous studies, we recognized and collected information of 71 compounds. The DrugCIPHER algorithm was applied to calculate the genome-wide target profiles of each JHT compound. The prediction accuracy for the major compounds ranged from 71% to 94% (Fig. 1A). To further compare the predicted target profiles across different compound types, we first categorized the 71 components of JHT into seven classes based on their chemical structure and calculated the proportion of each type (Fig. 1B). Alkaloids represented the largest group, accounting for 32% of all compounds, followed by organic acids at 23%, while phenylpropanoids were the least represented, comprising only 4%.

Additionally, the clustering results of the predicted target profiles for JHT components showed that compounds of the same type tended to cluster together (Fig. 1C). Most of the organic acids, flavonoids, and phenylpropanoids clustered into one group, suggesting that these three types of compounds may act in concert through similar biological processes or pathways. Among the four JHT herbs of *Melia toosendan* (MT, Chuanlianzi), *Corydalis yanhusuo* (CY, Yanhusuo), *Vladimiria souliei* (VS, Chuanmuxiang), and *Illicium dunnianum* (ID, Honghuabajiaoye), holistic targets of each herb were identified and analyzed (Fig. 1D). The statistical results indicate that

CY has 148 unique targets, while 95 targets are shared exclusively among ID, VS, and MT. Additionally, the number of unique targets for MT, VS, and ID are 26, 24, and 9, respectively. These findings suggest that despite differences in the chemical compositions of these herbs, they may still exert similar biological functions, aligning with the principles of network pharmacology, which emphasizes the multi-component, multi-target characteristics of interventions.

**JHT reduces inflammation levels in gastric tissue of CSG rats**

The experimental schedule and design are as illustrated in Fig. 2A. In control groups (C), the rats were administered normal saline via gavage and maintained under standard feeding conditions. For experimental groups, the rats underwent a comprehensive approach and were subsequently divided into three groups based on gavage treatment: the model group (M) receiving normal saline, the low-dose JHT group (D), and the high-dose JHT group (G).

Following the 72-day experimental period, H&E staining was conducted and diagnosed on gastric tissues from all groups to assess the extent of inflammation (Fig. 2B). It was found that the level of inflammatory cell infiltration in M group was higher than in groups C, D, and G. In M group, multiple inflammatory cells were observed infiltrating both the gastric mucosal epithelial layer to the muscularis mucosa and the muscularis mucosa to the submucosal layer. The pairwise t-test results demonstrated a significantly elevated inflammation score in the M group compared to the C group ($p < 0.001$), confirming successful model establishment (Fig. 2C). In addition, both the D group and G group exhibited significantly reduced inflammation scores compared to the M group ($p < 0.001$), with the G group showing an even lower mean inflammation score than the D group. This dose-dependent response indicated that an increased dose of JHT enhanced its therapeutic efficacy in reducing gastric inflammation associated with CSG.

**Network pharmacology analysis of JHT in the treatment of CSG**

Leveraging a previously established statistical model, the holistic targets of JHT and the composing herbs were identified. Subsequently, the holistic targets of JHT underwent KEGG and GO enrichment analysis. By integrating the KEGG enrichment results with documented pathological knowledge of CSG from literature and databases, 25 potential KEGG pathways through which JHT might intervene in CSG were identified. These KEGG pathways were categorized into four biological modules of CSG including immune regulation, cell growth and death, metabolism, and signal transduction, while GO enrichment analysis can further elucidate the specific biological process within these modules (Fig. 3).

Among them, gastritis was reported to be closely linked to immune regulation and signal transduction modules. Chronic inflammation in the gastric tissue triggers an immune response including the production of inflammatory cytokines such as IL-1β, IL-6, and TNF-α. These responses are regulated by signaling pathways such as TNF, NF-κB, and Toll-like receptors (TLRs), which activate immune cells and drive persistent inflammation (Wessler et al., 2017; Malfertheiner et al., 2023). Chronic gastritis is also closely associated with various cellular mechanisms, including apoptosis, necrosis, and senescence (Bertheloot et al., 2021). Additionally, chronic gastritis can induce metabolic disorders, with significant alterations in metabolic pathways (Takeoka et al., 2016; Yu et al., 2021).

Subsequently, we established a multi-layer biomolecular network based on the biological modules of CSG (Fig. 3A). It can be seen that the four herbs of JHT intervene in CSG through potential targets such as SOAT2, MC2R, IL1B, and PIK3CA. These interventions affect four biological modules, including immune regulation, cell growth and death, metabolism, and signal transduction. Leveraging RNA sequencing technology, our previous study (B. Wang et al., 2024a) demonstrated that JHT can intervene in CSG through regulating biological modules including immune regulation of pathways like Toll-like receptor signaling pathway, cell growth and death of pathways like Necroptosis, and signal transduction of pathways like MAPK and Sphingolipid signaling pathway. These findings are consistent with the results of network pharmacology analysis. However, the metabolic module-related mechanisms require further investigation and additional experiments.

Additionally, associations between these key modules and traditional effects were established using information from TCM theory, the SoFDA database (Zhang et al., 2022), and the Symmap database (Wu et al., 2019). Shugan Jieyu (SGJY) refers to regulating liver function, relieving liver Qi stagnation, and alleviating physical discomfort caused by emotional distress, such as stress, anxiety, or depression. This process is closely related to the signal transduction and immune modulation pathways. Liqi Huoxie (LQHX) works by regulating Qi, promoting blood circulation, and preventing Qi stagnation and blood stasis. It improves blood circulation and oxygen supply, directly supporting tissue repair and regeneration, reducing cell death, and promoting the growth of new cells. In addition, LQHX also accelerates metabolic processes in the body, particularly digestion and absorption in the gastrointestinal tract. Hewei Zhitong (HWZT) functions by regulating the stomach's digestive activity, relieving discomfort and pain in the stomach. Through the signal transduction pathway, it reduces the production of inflammatory mediators, which helps alleviate stomach pain and promotes the repair of gastric mucosa.

To further investigate the metabolic mechanisms through which JHT intervenes in CSG, we categorized the significantly enriched metabolic biological processes from the GO enrichment analysis into six metabolic categories: lipid, nucleotide, energy, carbohydrate, amino acid and other metabolism, based on the type and function of the metabolites (Fig. 3B). It can be found that the most prominent enrichment was observed in lipid metabolism, particularly in the regulation of lipid metabolic and biosynthetic processes. Phospholipid-related metabolic processes also represent a key mechanism through which JHT intervenes in CSG. In addition to lipid metabolism, nucleotide metabolism displayed significant enrichment, with key processes such as mRNA metabolic process being highly represented. Energy metabolism, specifically ATP and nitric oxide metabolic processes, also showed notable enrichment, suggesting their important role in the metabolism. These findings demonstrated the complexity of the potential mechanism, involving diverse metabolic processes. For further elucidation, metabolomic and microbiomics experiments were performed in this study.

**Metabolomic profiling of JHT in the treatment CSG**

We employed an untargeted metabolomics approach using UHPLC-Orbitrap-MS/MS to detect and assess the impact of JHT on the metabolic profile of CSG rats. Utilizing Wilcoxon rank-sum test, 20 differential metabolites were identified and a series of metabolomics analyses were performed on them (Fig. 4), where the abbreviations of these metabolites were shown in Table 2. The PCA results indicate that the G group shows a trend towards the C group, suggesting that high-dose JHT demonstrates an advantage in treating CSG (Fig. 4A). The heatmap illustrates the standardized abundance levels of differential metabolites across different groups (Fig. 4B), with phospholipids such as PS(19:0/0:0), PS(20:0/0:0), and PE(20:0/0:0), as well as citrulline, exhibiting significantly elevated levels in the model group. After JHT treatment, the abundance levels in the D and G groups were restored to levels similar to those in the control group. Additionally, a few metabolites, such as betaine, showed an opposite trend, with lower expression in the M group compared to the C and G groups.

The box plots provide a detailed representation of the standardized abundance of these five metabolites, along with the significance of differences between groups (Fig. 4C). The volcano plots offer a more intuitive visualization of the upregulation and downregulation of metabolite abundance, comparing the model group with the control group, as well as the high-dose JHT group with the model group (Fig. 4D). It can be observed that phospholipids and citrulline are upregulated in the model group and downregulated after treatment, whereas betaine shows the opposite trend. Studies have shown that the phospholipid levels in the gastric mucosa of patients with superficial gastritis increase, providing protection to the stomach (Matsuda et al., 2003).

To further analyze the metabolic mechanism of action of JHT in the treatment of CSG, a pathway enrichment analysis of the differential metabolites was conducted. The top 25 significantly enriched metabolic pathways (P value < 0.05) are shown in Fig. 4E. Among them, butyrate-induced histone acetylation is the most significantly enriched pathway. As an important short-chain fatty acid (SCFA), butyrate has been reported to be associated with immune regulation and inflammatory responses (He et al., 2020; Yu et al., 2024). In addition, reuptake of GABA, effects of nitric oxide, and lipid metabolism pathway also shows high levels of enrichment. γ-aminobutyric acid (GABA) is closely related to nervous system function and exhibits significant antidepressant effects in mouse models of depression (T Fuchs et al., 2017). The remaining two pathways were consistent with the results of network pharmacology analysis.

**Gut microbiota diversity analysis in rat fecal samples**

To assess the changes in gut microbiota before and after JHT intervention in CSG rats, we collected fecal samples and performed 16S rRNA sequencing (Fig. 5). The analysis results of PLS-DA show a clear distribution difference between the M group and the other groups, with the D and G groups gradually aligning with the C group along the COMP2 axis (Fig. 5A). This indicates that the drug intervention has a regulatory effect on the composition of the gut microbiota in rat fecal samples. The number of shared and unique species across multiple groups or samples can be used to identify species unique to the disease group, aiding in the discovery of biomarkers for building disease diagnostic models. In this study, the number of shared genera is 142, with the number of unique genera in each group shown in Fig. 5B.

The community barplot provides a visual overview of the changes in species across different groups and the structure of the microbial community (Fig. 5C). At the phylum level, Firmicutes have the highest proportion, followed by Bacteroidetes and Actinobacteriota, together accounting for over 95% of the total microbiota. The genus-level heatmap analysis shows the relative abundance changes of microbial communities across different groups (Fig. 5D). Furthermore, differential analysis was performed to identify significant microbial differences between groups, aiming to uncover key microorganisms associated with disease progression, treatment, or prognosis (Fig. 5E). At the genus level, the most significant differences (P value < 0.01) were observed in Bifidobacterium, Faecalibaculum, Dorea and so on, which belong to the lower-to-moderate abundance range within the overall microbial community. Notably, the abundance of Bifidobacterium and Faecalibaculum increased significantly after treatment, suggesting that these genera may play important roles in both disease states and therapeutic interventions.

Lastly, Lefse multi-level species differential analysis was used to identify microbes with the most significant contributions to group differences (Fig. 5F). In the healthy group, families such as Muribaculaceae and the Eubacterium coprostanoligenes group showed the highest abundance, indicating their potential role in maintaining gut homeostasis. Conversely, genera such as Rikenellaceae_RC9_gut_group, Lactococcus, and Butyricicoccus were most abundant in the model group, suggesting their association with disease onset and progression. Additionally, the increased abundance of Blautia, Bifidobacterium, Dubosiella, Faecalibaculum, and Turicibacter in the high-dose JHT group which indicates their potential role in treating CSG.

**Gut microbiota diversity analysis in rat cecal content samples**

16S rRNA sequencing was also applied on cecal content samples. Similar to the fecal microbiota analysis, we conducted a supplementary diversity analysis of the gut microbiota in cecal content samples (Fig. 6).

After applying supervised analysis methods, the control and model groups were clearly distinguishable, with the low-dose JHT group showing a slight shift towards the control group, though the regulatory effect was not prominent. In contrast, the high-dose JHT group exhibited significant regulatory effects along the COMP2 axis (Fig. 6A). All groups shared 124 genera, while each group also had unique genera, indicating differences in microbial composition among the experimental groups (Fig. 6B). At the phylum level, Firmicutes had the highest relative abundance across all groups, followed by Actinobacteriota, with the two phyla accounting for over 90% of the total microbiota. In comparison, the proportion of Bacteroidetes was lower than that observed in fecal samples (Fig. 6C). The heatmap provides an overview of the microbial genera present across the disease, healthy, and treatment groups, along with their relative abundances (Fig. 6D).

Statistical analysis of microbial abundance revealed significant differences among groups. Genera such as Dubosiella, Faecalibaculum, and Bifidobacterium exhibited significant increases in abundance following treatment. In contrast, genera such as Adlercreutzia and Enterorhabdus showed reduced abundance after treatment, shifting towards levels similar to the control group (Fig. 6E). The Lefse multi-level species differential analysis (Fig. 6F) identified several key genera with significant group-specific enrichment. In the control group, genera with higher abundances included norank_f__Ruminococcaceae, Bilophila, and Eubacterium_ruminantium_group. In the model group, Adlercreutzia, Ruminiclostridium, and Enterorhabdus were more abundant. The low-dose JHT group was enriched with Butyrivibrio, Faecalibaculum, and norank_f__norank_o__RF39, while the high-dose JHT group showed increased abundances of Dubosiella, Bifidobacterium, and Mucispirillum. These findings highlight and further support the potential regulatory effects of JHT on the gut microbiota, providing a foundation for deeper understanding of the mechanisms underlying drug intervention.

**Discussion**

Following the holistic philosophy, this study establishes a comprehensive analytical framework based on animal study, network pharmacology, metabolomics, and gut microbiota to systematically explore the mechanism of JHT in treating CSG in rats, further elucidating the potential metabolic-related biological effects of JHT on CSG. JHT is commonly used to treat liver-stomach disharmony syndrome in chronic superficial gastritis. The animal study performed in this study demonstrated that JHT significantly reduced the inflammatory levels in the gastric tissues of CSG rats. The network pharmacology analysis suggested that metabolomic and microbiomic experiments were needed to further elucidate and validate. A key aspect of the study is the discovery of correlations between metabolites and gut microbiota through untargeted metabolomics analysis and gut microbiota profiling, validating the results of the network pharmacology analysis. Notably, following JHT administration, the abundance of metabolites such as betaine, citrulline, and phospholipids (including PE and PS) exhibited significant changes. Additionally, the abundance of bacterial genera such as Bifidobacterium and Faecalibaculum increased markedly, indicating that JHT may exert its anti-CSG effects by regulating the interplay between metabolites and gut microbiota (Fig. 7). These findings provide new insights into the mechanisms underlying JHT's therapeutic effects on CSG, supporting the theoretical foundations of TCM in the intervention of chronic gastritis.

Using a rat model of CSG conducted in the previous study (B. Wang et al., 2024a), it was shown that the inflammation scores of the model group, assessed by H&E staining, were significantly higher than those of the control group, and the inflammation scores were significantly reduced after the intervention of JHT, which was consistent with the results of the previous clinical studies (Zhang et al., 1998). The results provided evidence to the efficacy of JHT in treating CSG.

To explore the potential mechanisms of JHT, network pharmacology analysis was conducted. This involved the integration of compound targets prediction, holistic targets calculation, enrichment analysis, and multi-layer biomolecular network construction. The results revealed that JHT exerts a synergistic therapeutic effect on CSG by modulating pathways across multiple modules: immune regulation pathways such as the Toll-like receptor pathway, cell growth and death pathways including apoptosis, signal transduction pathways like TNF and MAPK signaling, and metabolism-related pathways.

The results of network pharmacology analysis aligned with the findings of several previous studies. Malfertheiner et al. reported that the immune response triggered by chronic gastritis is regulated by signaling pathways such as TNF and Toll-like receptor pathways (Malfertheiner et al., 2023). Moreover, chronic gastritis is closely associated with various cellular mechanisms, including apoptosis, necrosis, and senescence (Bertheloot et al., 2021). Metabolic pathways, particularly lipid metabolism, are also known to induce metabolic disorders in patients with chronic gastritis (Takeoka et al., 2016). To further investigate metabolism-related biological processes, GO enrichment analysis highlighted JHT's significant regulatory effects on phospholipid metabolism, ATP metabolism, and nitric oxide metabolism and biosynthesis in CSG rats. These insights provide valuable guidance for subsequent experimental studies, shedding light on the critical metabolic processes involved in JHT's intervention against CSG.

Plasma metabolomic study analyzed the difference in the levels of plasma metabolites among normal, model, and treatment groups, serving as systemic indicators of the holistic conditions. Building on the metabolic profile of JHT identified through UPLC-Q/TOF-MS analysis, where 38 prototype compounds and 125 metabolites were detected (Liu et al., 2021), our study employed untargeted metabolomics to investigate the effects of JHT intervention in CSG rats. After data normalization, 20 differential metabolites including betaine, citrulline, PE, and PS were identified. Pathway enrichment analysis revealed that these metabolites may exert therapeutic effects by modulating SCFA-related pathways, nitric oxide synthesis, and lipid metabolism pathways.

Phospholipid levels were significantly increased in the model group and decreased after treatment. This aligns with previous research showing elevated phospholipid levels in the gastric mucosa of CSG patients, which also indicated that phospholipids play a role in enhancing the gastric mucosal barrier (Matsuda et al., 2003). The reduction in phospholipid levels after treatment may reflect improved inflammation, suggesting phospholipid levels could be used to assess the severity of gastric inflammation. Interestingly, citrulline levels followed a similar pattern, increasing in the model group and decreasing post-treatment. As a byproduct of nitric oxide (NO) synthesis, citrulline indicates the activity of nitric oxide synthase (NOS) and serves as a marker of NO levels (Keilhoff et al., 2000). Elevated inducible NOS (iNOS) levels have been reported in the gastric tissue of CSG patients (Pignatelli et al., 1998). These findings align with our network pharmacology predictions, supporting JHT's ability to regulate NO metabolism and biosynthesis. Thus, in this study, citrulline levels provide an indirect measure of NO production, helping to assess the severity of gastric inflammation in CSG rats.

Furthermore, by leveraging metabolomics, we observed a significant decrease in betaine levels in the model group, which was restored post-treatment. Betaine, an amino acid derivative (Dobrijević et al., 2023), enriches gut microbiota and strengthens the intestinal epithelium, thus supporting intestinal health (Arumugam et al., 2021). SCFAs, once produced in the gut, can enter the peripheral bloodstream, modulating functions in other organs and playing a pivotal role in inflammation and lipid metabolism (He et al., 2020). Kan et al. provided evidence linking SCFA levels with CSG development (Kan et al., 2020). Additionally, SCFAs act as substrates for mitochondrial ATP synthesis, aligning with the network pharmacology prediction of JHT's role in ATP metabolic processes (Panagia et al., 2020; Tan et al., 2023).

The gut microbiota is closely linked to other organs such as the brain and lungs, they work together to maintain the homeostasis of the body (Cryan et al., 2019; Dang and Marsland, 2019). Previous studies revealed that gut microbiota influenced immune regulation and metabolism in gastritis (Man, 2018; Malfertheiner et al., 2022). Combining gut microbial analysis and metabolomics helped uncovering the mechanism of drug intervention from a systemic perspective. In this study, we both collected the fecal and cecal content samples and subjected them to the analyses of 16S rRNA sequencing. The results demonstrated that Faecalibaculum and Bifidobacterium had low abundance in control and model groups but significantly increased after treatment, suggesting their important role in JHT's therapeutic effect on CSG. It has been shown that Faecalibaculum can produce SCFAs to inhibit the growth of intestinal tumors (Zagato et al., 2020), while Bifidobacterium's fermentation products support the growth of other SCFA-producing bacteria (Mann et al., 2024). Therefore, we speculate that JHT intervention raises betaine levels, enriches gut microbiota, and enhances the intestinal barrier. Increased Faecalibaculum and Bifidobacterium levels promote SCFA production, which, through circulation, may regulate gastric inflammation.

Nevertheless, this study has several limitations that warrant further investigation. Firstly, we did not explore the molecular mechanisms at the gene and protein levels, which could be addressed through transcriptomics and proteomics analyses to uncover molecular pathways. Secondly, the untargeted metabolomics analysis relied solely on peripheral blood samples. Expanding sample types to urine or feces could provide a more comprehensive understanding of the metabolic mechanisms underlying the drug intervention. Lastly, while 16S rRNA sequencing is effective for microbial diversity analysis, it has limitations such as low species resolution and primer bias. Future studies should incorporate metagenomics to identify specific taxa and microbial communities more accurately. Addressing these challenges will enhance our understanding of CSG pathogenesis and reveal the metabolic and microbial mechanisms through which JHT exerts its therapeutic effects.

**Conclusion**

This study confirmed the anti-inflammatory effects of JHT on gastric tissues in CSG rats and explored its underlying mechanisms. From the perspective of holistic concept, we integrated network pharmacology, untargeted metabolomics, and gut microbiota analysis and demonstrated that JHT exerted its therapeutic effects by modulating the abundance of metabolites including betaine, citrulline, and phospholipids, as well as altering the abundance of Faecalibaculum and Bifidobacterium. Our findings suggested that JHT might enhance SCFA production by indirectly influencing microbial abundance including Faecalibaculum and Bifidobacterium through betaine, which helped alleviate gastritis. However, due to the complexity of TCM formulas, further research is needed to clarify the specific sources and pathways of key metabolites and to explore the intermediate mechanisms by which gut microbiota exerts its effects. Overall, this study highlighted JHT's therapeutic potential for CSG by uncovering novel mechanisms of inflammation reduction.

## CRediT authorship contribution statement

Tingyu Zhang: Conceptualization, Formal analysis, Investigation, Visualization, Writing – original draft. Lihao Xiao: Investigation, Data curation, Resources, Validation, Visualization, Writing – review & editing. Yun Liu: Conceptualization, Formal analysis, Visualization. Chayanis Sutcharitchan: Investigation, Visualization, Writing – review & editing. Qingyuan Liu: Investigation, Visualization, Writing – review & editing. Xiaoxue Fan: Data curation, Resources, Validation. Jian Feng: Data curation, Resources, Validation. Huifang Gao: Data curation, Resources, Validation. Tong Zhang: Project administration, Supervision, Validation. Shao Li: Project administration, Supervision, Visualization, Writing – review & editing.

## Conflicts of Interest

The authors declare that there are no known conflicts of interest associated with this publication and there has been no significant financial support for this work that could have influenced its outcome.

**Acknowledgments**

Some of the elements in Fig. 7 are obtained from Freepik.

**Funding**

This work was supported by grants from the National Natural Science Foundation of China (Grant No.: T2341008), Intelligent and Precise Research on TCM for Spleen and Stomach Diseases (20233930063), and Basic Research Program Natural Science Fund - Frontier Leading Technology Basic Research Special Project of Jiangsu Province (No. BK20232014).

**Table**

Table 1. Software and databases for gut microbiota analysis

| Software/database | Version | Purpose | Link |
|---|---|---|---|
| Qiime | 1.9.1 | Sequence deduplication and filtering, sequence classification and annotation, beta diversity distance calculation | http://qiime.org/install/index.html |
| Uparse | 7.0.109 | OTU clustering | http://www.drive5.com/uparse/ |
| Usearch | 7.0 | OTU statistics | http://www.drive5.com/usearch/ |
| Mothur | 1.30.2 | Alpha diversity analysis | https://www.mothur.org/wiki/Download_mothur |
| MAFFT | 7.2 | Multiple sequence alignment | https://mafft.cbrc.jp/alignment/software/ |
| IQ-TREE | 1.6.8 | Phylogenetic tree construction | http://www.iqtree.org |

Table 2. Abbreviations for differential metabolites

| Compound name | Abbreviation |
|---|---|
| 1-(2-methoxy-eicosanyl)-sn-glycero-3-phosphoserine | M-Eico-PS |
| 2-(8-Hydroxy-4a,8-dimethyldecahydro-2-naphthalenyl)acrylic acid | Hydroxy-Naph-AA |
| 2-Amino-1-methyl-6-phenylimidazo(4,5-b)pyridine | Amino-Methyl-PP |
| 2-Dodecylbenzenesulfonic acid | Dodecyl-BSA |
| 2-Oxo-delta3-4_5_5-trimethylcyclopentenylacetate | Oxo-Trimethyl-Cyclo |
| 4-Dodecylbenzenesulfonic acid | Dodecyl-BSA-4 |
| 4-Undecylbenzenesulfonic acid | Undecyl-BSA-4 |
| 6-Amino-3-methyl-1-phenyl-1H-pyrazolo[3,4-b]pyridine-5-carboxamide | Amino-Pyrazolo |
| Betaine | Betaine |
| Diethylene glycol | Diethylene-Glycol |
| Eugenolmethylether | Eugenol-ME |
| L(+)-Citrulline;2-Amino-5-uredovalerate | Citrulline |
| LSD-d3 | LSD-d3 |
| LT9970000 | LT9970000 |
| OKHdiA-PE | OKHdiA-PE |
| Ostruthin | Ostruthin |
| PE(20:0/0:0) | PE(20:0/0:0) |
| Pregna-4_9(11)-diene-3_20-dione | Pregna-Diene |
| PS(19:0/0:0) | PS(19:0/0:0) |
| PS(21:0/0:0) | PS(21:0/0:0) |

# Figure

Fig. 1. Target prediction, validation and comparison among different compounds set in JHT. (A) Target prediction accuracy of major compounds in JHT. (B) Proportion of seven chemical components types in JHT. (C) Hierarchical clustering of JHT compounds. (D) Upset plot for holistic targets sets of four herbs in JHT.

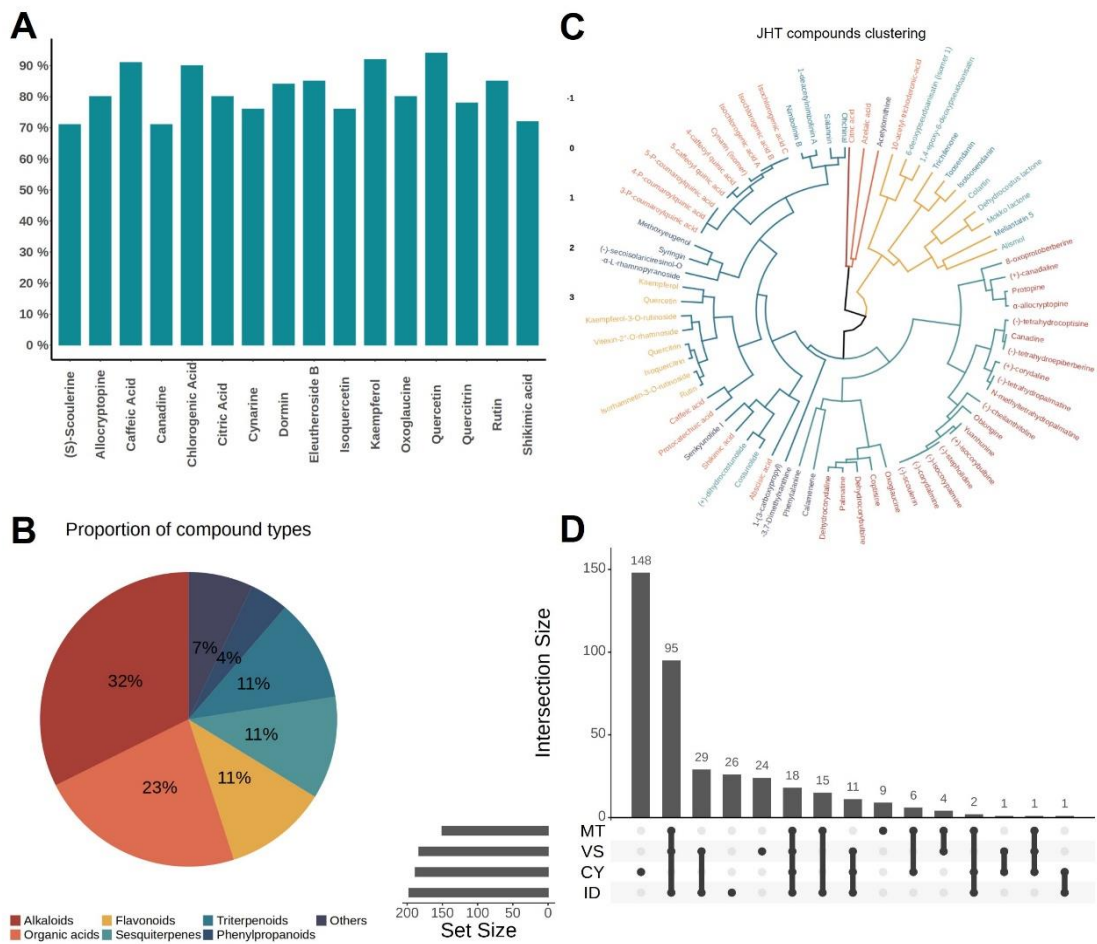

Fig. 2. JHT improves the inflammation in gastric tissues of CSG rats. (A) Experimental design and schedule in treatment of CSG using JHT. (B) Representative photomicrographs among four groups (Magnification 100×, Scale bar=250μm). (C) Statistical tests for inflammation scores of gastric tissues in different groups, with mean and standard deviation above the bars. *p<0.05, **p<0.01 and ***p<0.001 as compared between two groups.

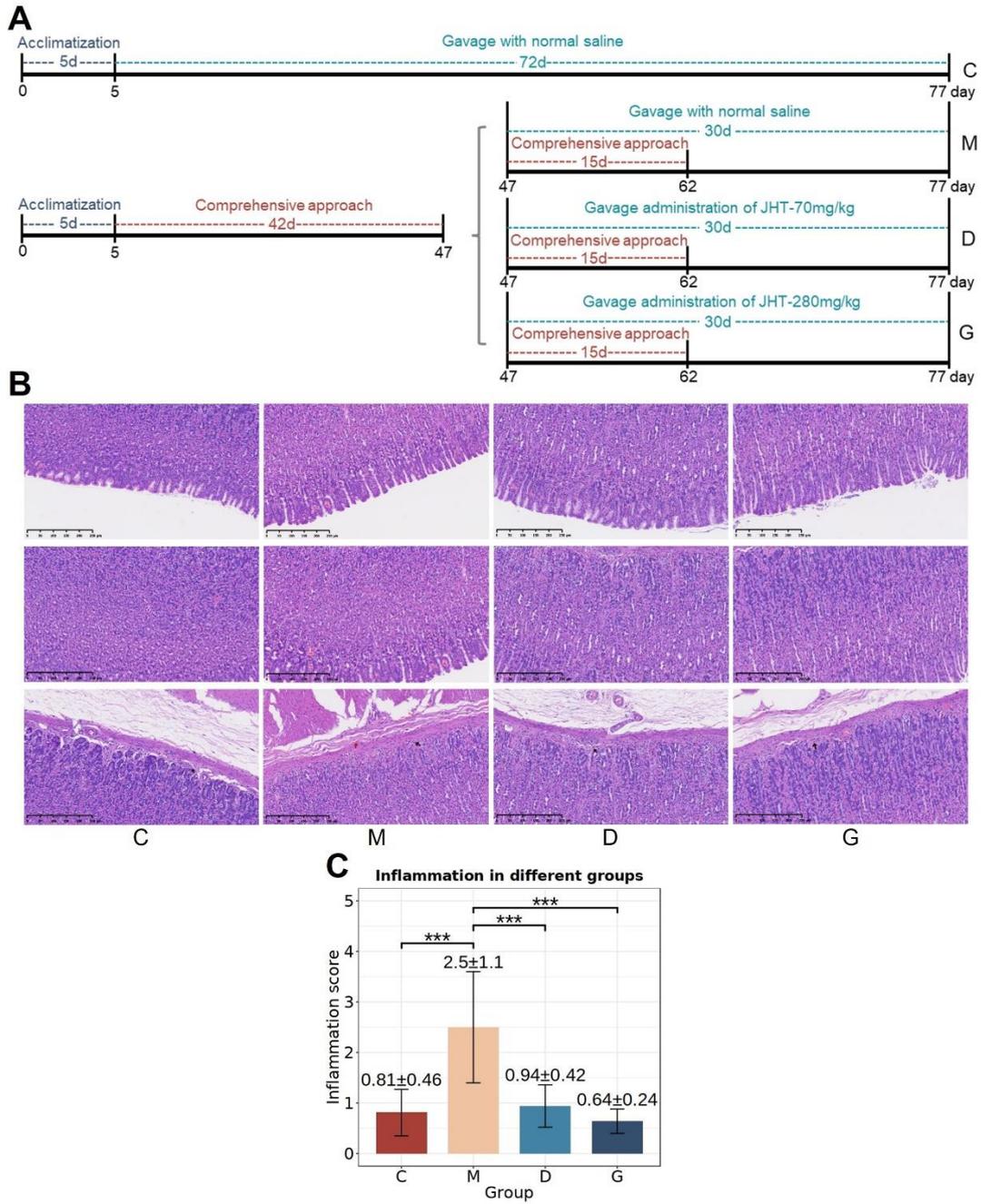

Fig. 3. Network pharmacology analysis of JHT in modules of CSG. (A) Multi-layer biomolecular network representing the mechanism of action of JHT in the treatment of CSG. SGJY Shugan Jieyu, LQHX Liqi Huoxie, HWZT Hewei Zhitong. (B) Enrichment plot of GO biological processes in metabolism of JHT in the treatment of CSG.

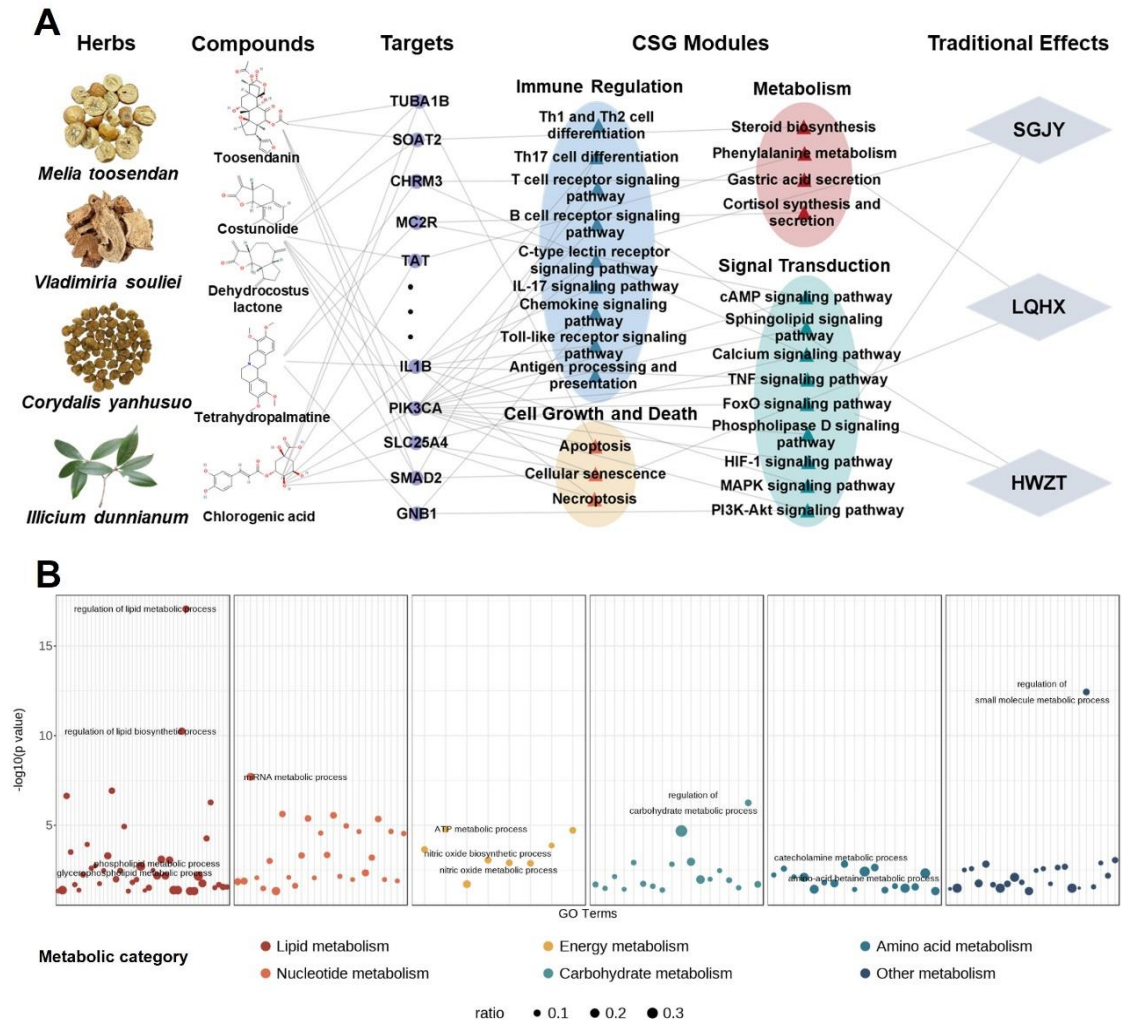

Fig. 4. Comprehensive metabolomics and pathway enrichment analysis of JHT's therapeutic effects in CSG. (A) PCA Plot showing the distribution of samples from each group on the two main components. (B) Heatmap demonstrating the standardized abundance of differential metabolites between groups. (C) Volcano plots revealing the standardized abundance of 5 key differentially expressed metabolites varied between different groups. Not significant (NS), *p<0.05, **p<0.01 and ***p<0.001 as compared between two groups. (D) Boxplots displaying the upregulation and downregulation of metabolites between two groups. (E) Pathway enrichment analysis representing 25 significantly enriched metabolic pathways after JHT treatment for CSG, sorted by enrichment ratio.

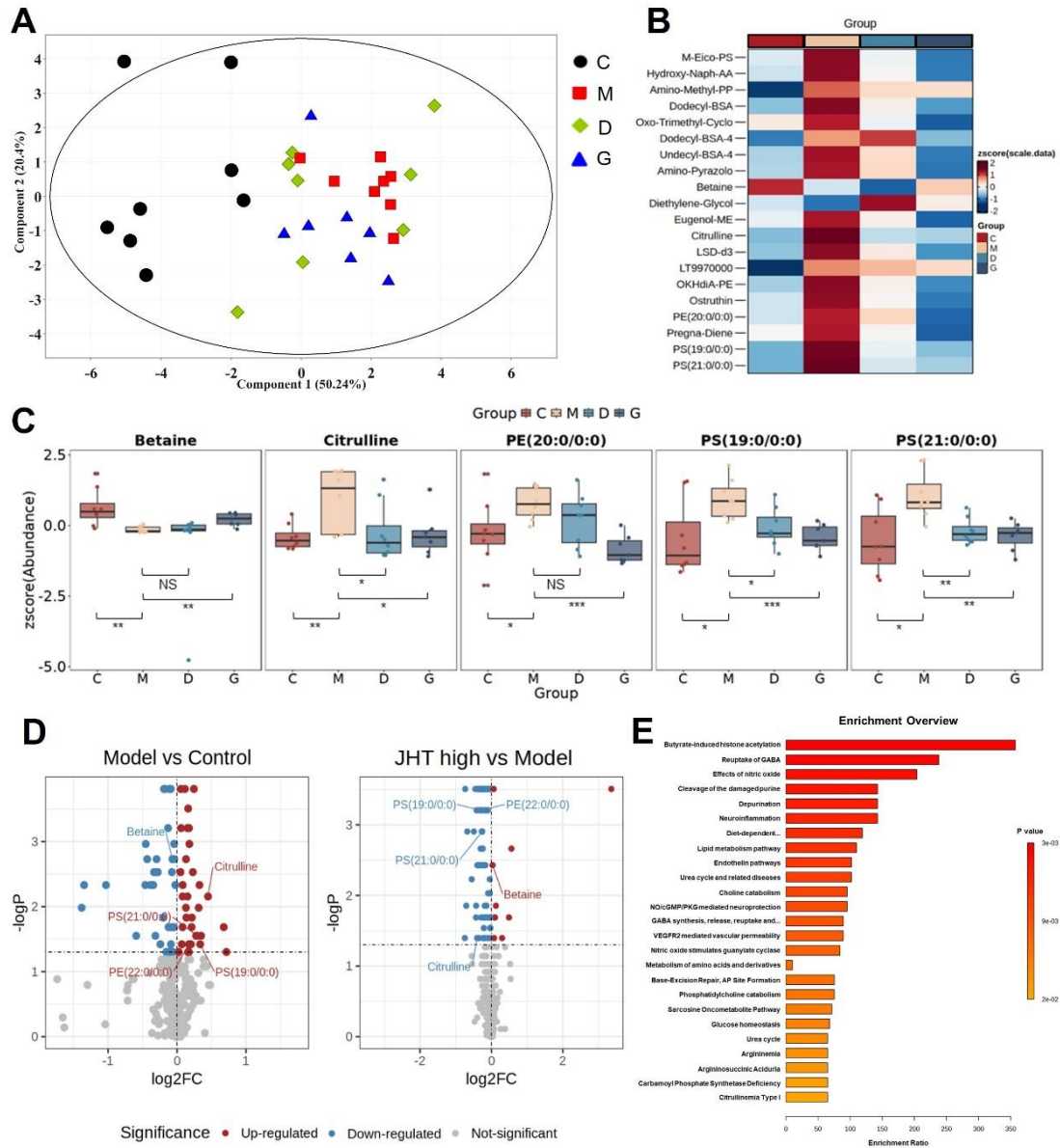

Fig. 5. Diversity analysis of gut microbiota in rat fecal samples across experimental groups. (A) PLS-DA plot showing the distribution of samples at the OTU level along the two main components. (B) Venn diagram depicting the number of shared and unique OTUs among the experimental groups. (C) Community barplot showing the relative abundance of microbial communities at the phylum level across different groups. (D) Genus-Level heatmap illustrating the relative abundance of microbial genera across groups. (E) Kruskal-Wallis H Test Bar Plot showing significant differences in microbial taxa among groups, with distinct taxa enriched in specific groups. *p<0.05, **p<0.01. (F) Cladogram representing the phylogenetic relationships of microbial taxa, with nodes and branches color-coded to indicate group-specific enrichment.

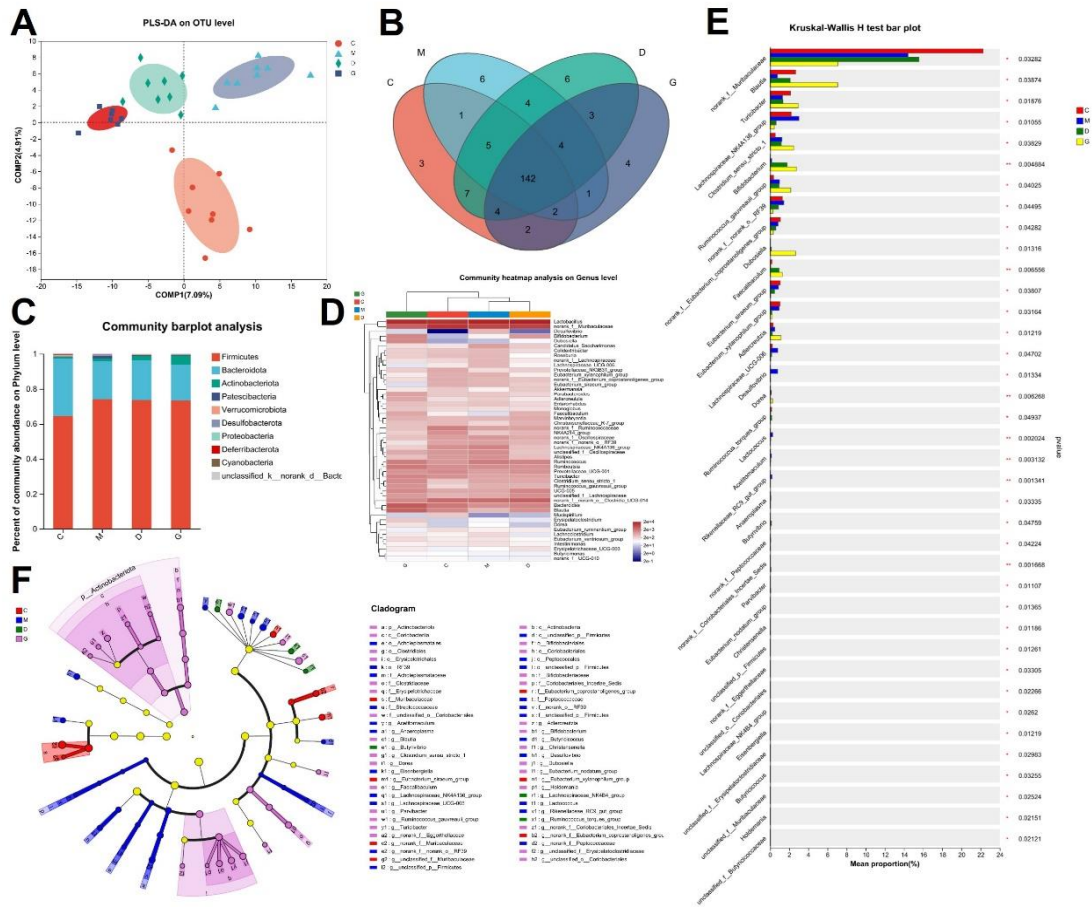

Fig. 6. Diversity analysis of gut microbiota in rat cecal content samples across experimental groups. (A) PLS-DA plot depicting the distribution of samples at the OTU level along the two main components. (B) Venn diagram showing the number of OTU intersection and difference across different groups. (C) Community barplot displaying the relative abundance of microbial communities at the phylum level among the experimental groups. (D) Genus-Level heatmap illustrating the relative abundance of microbial genera among groups. (E) Kruskal-Wallis H test barplot demonstrating significant differences in microbial abundance across different groups. *p<0.05, **p<0.01, ***p<0.01. (F) Cladogram showing the significant enrichment of microbial communities between different groups, with different colors representing bacterial communities in different groups.

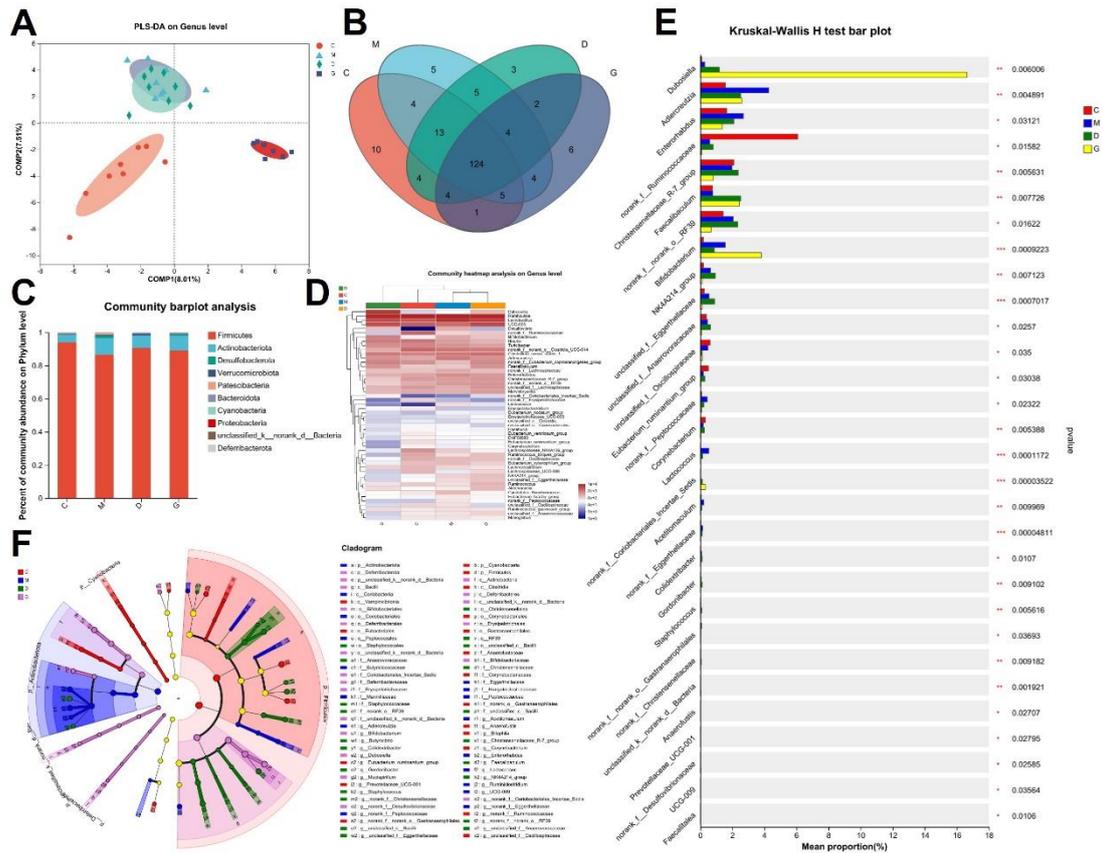

Fig. 7. Mechanistic insights into JHT's therapeutic effects on CSG rats via gut microbiota and metabolite regulation.

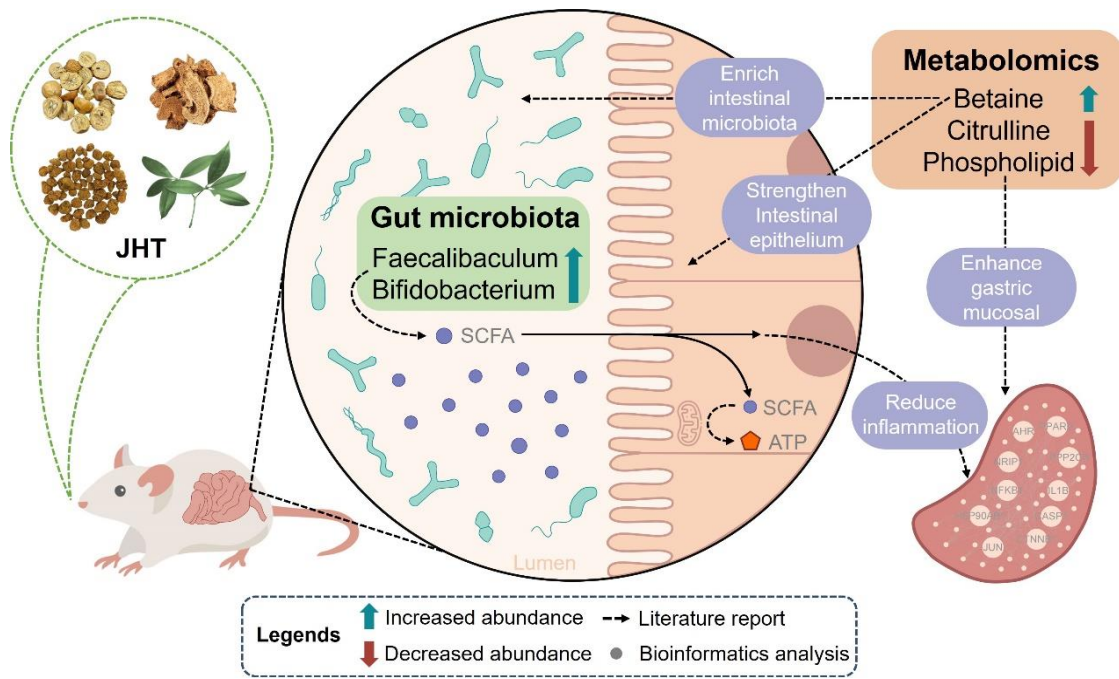